\pdfoutput=1
\documentclass[preprint]{acmart}

\renewcommand\footnotetextcopyrightpermission[1]{} 
\AtBeginDocument{%
  \providecommand\BibTeX{{%
    \normalfont B\kern-0.5em{\scshape i\kern-0.25em b}\kern-0.8em\TeX}}}

\setcopyright{none}

\usepackage{balance}       
\usepackage[T1]{fontenc}   
\usepackage{color}
\usepackage{booktabs}
\usepackage{textcomp}
\usepackage{adjustbox}
\usepackage{caption,subcaption,graphicx}

\usepackage{graphicx} 
\usepackage{amsmath}
\usepackage{amsfonts} 
\usepackage{times}  
\usepackage{helvet} 
\usepackage{courier}  
\usepackage{comment}
\usepackage{multirow}
\usepackage{adjustbox}
\usepackage{caption,subcaption,graphicx}

\usepackage{algorithm}
\usepackage{algorithmic}
\usepackage{hhline}

\usepackage{flafter} 
\usepackage{eucal}

\usepackage{microtype}        
\usepackage{ccicons}          

\newcommand*{\origrightarrow}{}

\renewcommand*{\textrightarrow}{\fontfamily{cmr}\selectfont\origrightarrow}

\begin{document}

\title{Opportunities of a Machine Learning-based Decision Support System for Stroke Rehabilitation Assessment}

\author{Min Hun Lee, Daniel P. Siewiorek, Asim Smailagic}
\email{{minhunl, dps, asim}@cs.cmu.edu}
\affiliation{%
  \institution{Carnegie Mellon University}
  \state{PA, USA}
}

\author{Alexandre Bernardino}
\affiliation{%
  \institution{Instituto Superior T{\'e}cnico}
  \city{Lisbon, Portugal}}
\email{alex@isr.tecnico.ulisboa.pt}

\author{Sergi Berm{\'u}dez i Badia}
\affiliation{%
  \institution{Madeira Interactive Technology Institute}
  \city{Funchal, Portugal}}
  \email{sergi.bermudez@m-iti.org}

\renewcommand{\shortauthors}{Lee, et al.}

\begin{abstract}
Rehabilitation assessment is critical to determine an adequate intervention for a patient. However, the current practices of assessment mainly rely on therapist's experience, and assessment is infrequently executed due to the limited availability of a therapist. In this paper, we identified the needs of therapists to assess patient's functional abilities (e.g. alternative perspective on assessment with quantitative information on patient's exercise motions). As a result, we developed an intelligent decision support system that can identify salient features of assessment using reinforcement learning to assess the quality of motion and summarize patient specific analysis. We evaluated this system with seven therapists using the dataset from 15 patient performing three exercises. The evaluation demonstrates that our system is preferred over a traditional system without analysis while presenting more useful information and significantly increasing the agreement over therapists' evaluation from 0.6600 to 0.7108 F1-scores ($p <0.05$). We discuss the importance of presenting contextually relevant and salient information and adaptation to develop a human and machine collaborative decision making system.
\end{abstract}



\maketitle

\section{Introduction}
Assessment of physical rehabilitation exercises is an essential process to determine an appropriate clinical intervention for a patient with musculoskeletal and neurological disorders (e.g. stroke) \cite{lang2013assessment}. However, this process relies on therapist's experience \cite{smith2008mispredictions} and is infrequently performed due to the limited availability of a therapist. Researchers have explored the possibility of computer-assisted decision support tools \cite{berner2007clinical} that can monitor and assess chronic diseases using sensor and machine learning technologies \cite{webster2014systematic}. 

\begin{figure}[htp!]
\centering
\begin{subfigure}{.25\columnwidth}
\centering
  \includegraphics[height=.7\columnwidth]{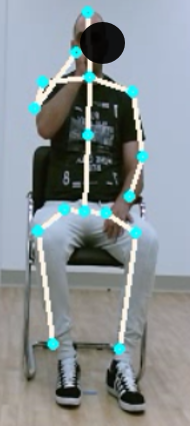}
  \caption{}
  \label{fig:sample-compensation-normal-p11}
\end{subfigure}\hfill  
\begin{subfigure}{.25\columnwidth}
\centering
  \includegraphics[height=.7\columnwidth]{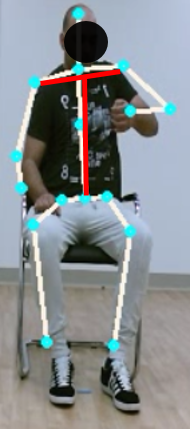}
  \caption{}
  \label{fig:sample-compensation-abnormal-p11}
\end{subfigure}\hfill  
\begin{subfigure}{.25\columnwidth}
\centering
  \includegraphics[height=.7\columnwidth]{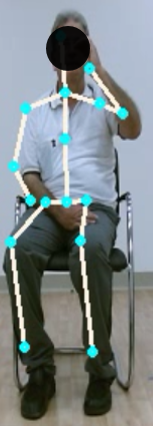}
  \caption{}
  \label{fig:sample-compensation-normal-p14}
\end{subfigure}\hfill  
\begin{subfigure}{.25\columnwidth}
  \centering
  \includegraphics[height=.7\columnwidth]{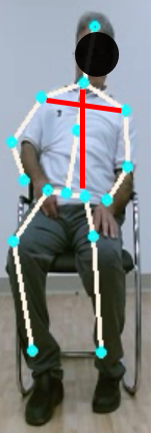}
  \caption{}
  \label{fig:sample-compensation-abnormal-p14}
\end{subfigure}
\caption{Two patients performing the \textit{`Bring a Hand to Mouth'} exercise with different compensated joints: (a) unaffected and (b) affected motions of patient 11 (elevated shoulder and trunk rotation). (c) unaffected and (d) affected motions of patient 14 (elevated shoulder and leaning backward).}~\label{fig:sample-compensation}. 
\end{figure}

For instance, a Support Vector Machine (SVM) classifier is applied to distinguish mild and severe symptoms of four Parkinson's patients \cite{das2011quantitative}. Neural Networks are utilized to quantify the quality of stroke rehabilitation exercises \cite{lee2019learning}. These approaches process complex sensor data to automatically extract a meaningful function, machine learning model to classify the quality of motion. However, it is challenging to derive a model that can perfectly replicate therapist's assessment given patient's diverse physical characteristics. For example, two patients could have different ways of incorrectly performing an exercise (Figure \ref{fig:sample-compensation}). Thus, a model can incorrectly predict a new patient's exercise motion with compensated joints that is not present in the dataset of a system. If a model with complex algorithms cannot explain why it provides different assessment \cite{gunning2017explainable}, therapists could lose trust in the model and abandon it even if it provides valuable predictions in other cases \cite{khairat2018reasons}.

\begin{figure} [htp]
\centering
  \includegraphics[width=0.5\columnwidth]{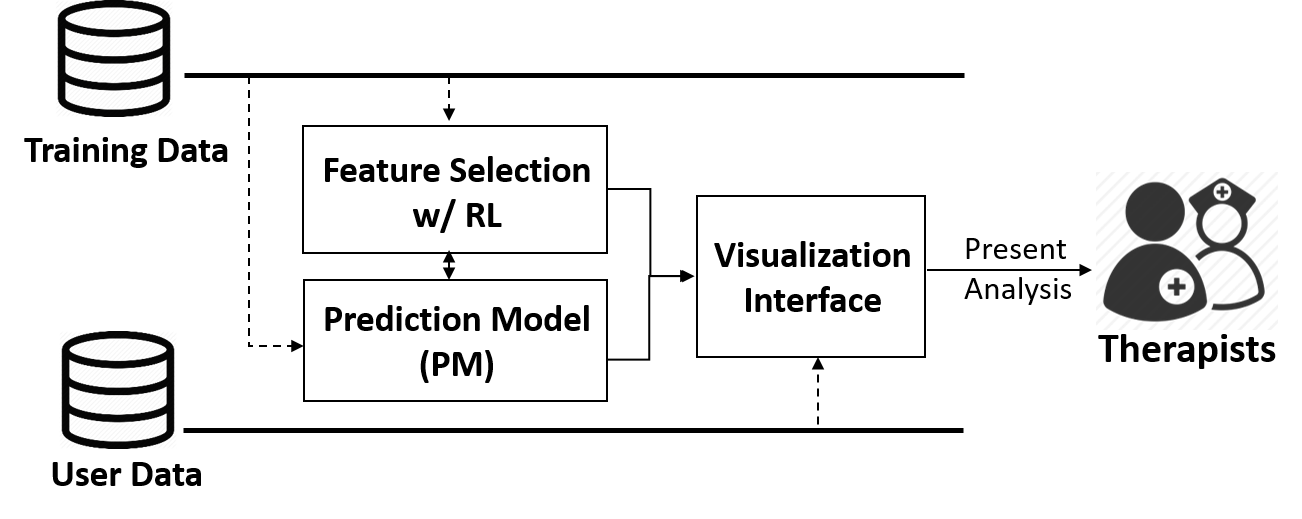}
  \caption{Flow diagram of an intelligent decision support system}~\label{fig:overall-arch}
\end{figure}

\begin{figure*}[htp!]
\centering 
\begin{subfigure}[t]{.24\textwidth}
\centering
  \includegraphics[width=0.93\columnwidth]{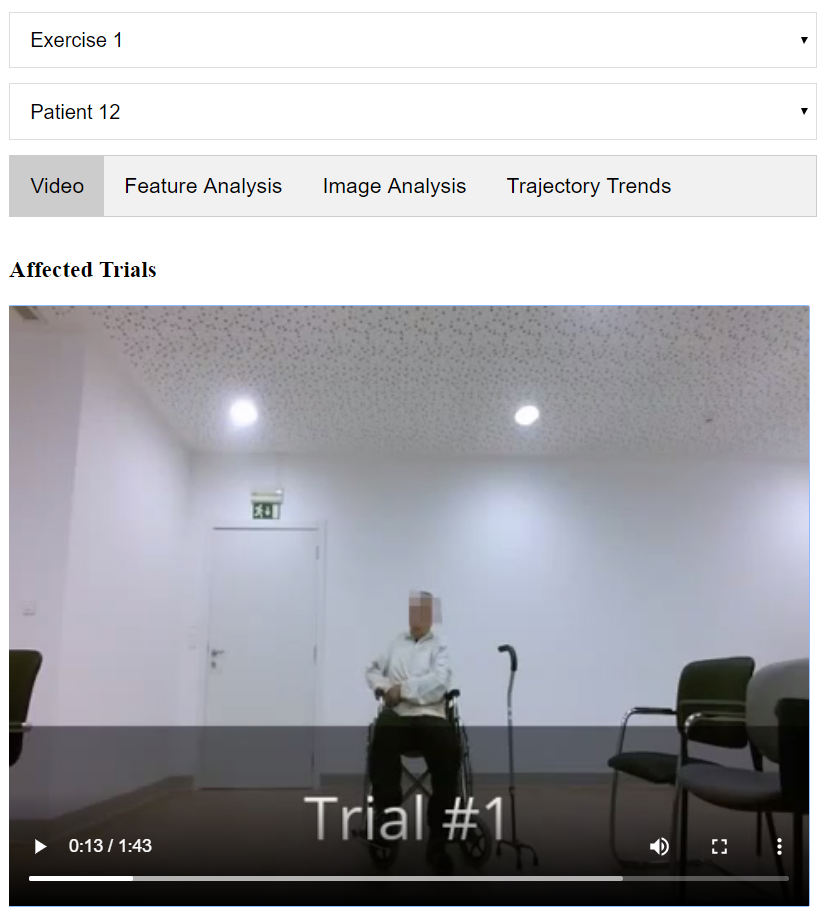}
  \caption{}
  \label{fig:web-video}
\end{subfigure}\hfill
\begin{subfigure}[t]{.24\textwidth}
  \centering
  \includegraphics[width=1.0\columnwidth]{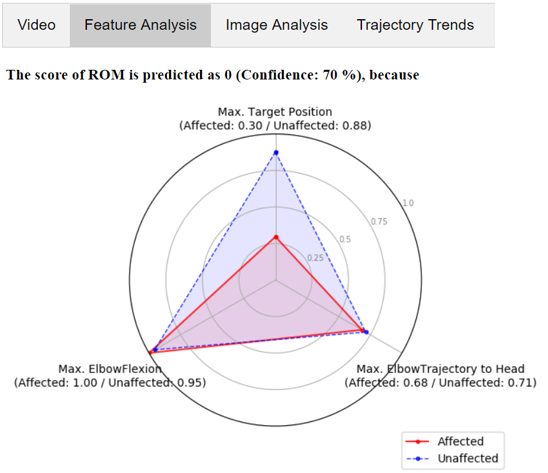}
  \caption{}
  \label{fig:web-feat}
\end{subfigure}
\begin{subfigure}[t]{.24\textwidth}
\centering
  \includegraphics[width=1.0\columnwidth]{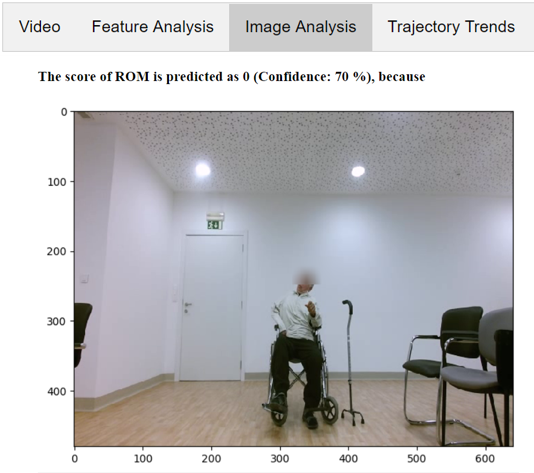}
  \caption{}
  \label{fig:web-img}
\end{subfigure}\hfill
\begin{subfigure}[t]{.24\textwidth}
  \centering
  \includegraphics[width=1.0\columnwidth]{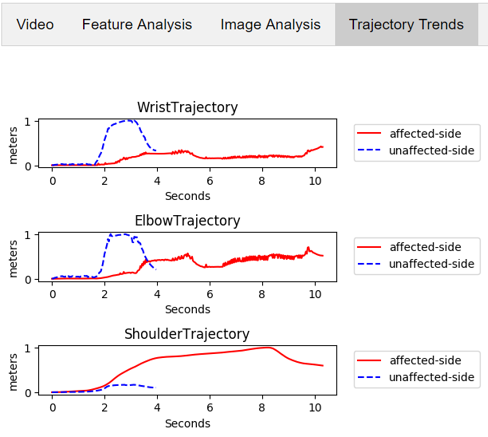}
  \caption{}
  \label{fig:web-traj}
\end{subfigure}
    \caption{The visualization interface of the proposed system that presents (a) a video of patient's exercise motions and a predicted performance score with (b) feature analysis between unaffected and affected side, (c) images of salient frames, and (d) trajectory trends between unaffected and affected side.}   \label{fig:web-interface}
\end{figure*}

In this paper, we implement and evaluate an intelligent decision support system (Figure \ref{fig:overall-arch}). This system utilizes training data of all patients except a patient for testing to learn models to identify salient features of assessment using reinforcement learning and predict the quality of motion with Neural Networks. Using identified salient features and held-out user data, patient's unaffected motions, this system can provide user-specific analysis with a visualization interface. 
This system empowers therapists to understand patient's performance with 1) feature analysis with kinematic measurements, 2) images of salient frames, and 3) trajectory trends (Figure \ref{fig:web-interface}). While considerable prior work demonstrates the feasibility of assessing the quality of motion \cite{webster2014systematic} and focuses on improving the accuracy of a model \cite{lee2019learning}, there is a lack of systematic evaluations on such technologies.

We conducted the field studies with therapists to identify what capabilities they need and implemented an intelligent decision support system for assessing stroke rehabilitation exercises with the exercise dataset (15 post-stroke patients performing three upper-limb exercises). This system presents the predicted scores of three performance components (i.e. \textit{`Range of Motion', `Smoothness', `Compensation'}) with user-specific analysis as explanations of the predictions: feature analysis, images of salient frames, and joint trajectories (Figure \ref{fig:web-interface}). We performed a user study with seven therapists from four rehabilitation hospitals to investigate how therapists use this system and how it can affect therapist's decision making on assessing patient's exercise performance. Results show that our system enables therapists to validate their assessment with quantitative, user-specific analysis, which increases user trust and utility of a system. In addition, our system assists therapists to achieve significantly higher agreement on their assessment (0.71 average F1-scores) than a traditional system without analysis (0.66 average F1-scores) ($p < 0.05$).

This paper makes the following contributions:

\begin{itemize}
    \item enumerate needs of therapists during assessing rehabilitation exercises
    \item present the design and implementation of an intelligent decision support system for stroke rehabilitation assessment that can identify salient features using reinforcement learning to predict the quality of motion and generate user-specific analysis
    \item describe the quantitative and qualitative evaluation of our system with seven therapists from four rehabilitation hospitals and pose this system as an approach to support consistent assessment
\end{itemize}

\section{Related Work}\label{sect:related}
\subsection{Current Practices of Physical Rehabilitation}
Patients with musculoskeletal and neurological disorders (e.g. stroke) require a rehabilitation program over several months to prevent disability and improve their functional abilities. Performing task-oriented exercises is one of the effective ways for post-stroke survivors to improve functional ability and lower a chance of having recurrent stroke \cite{rensink2009task}. During a rehabilitation program, therapists first diagnose the condition of a patient with various methods (e.g. analyzing patient's history, conducting tests, or analyzing measurements) and determine in-home interventions. In the follow-up visits, therapists discuss patient's progress or periodically evaluate treatment outcomes to modify interventions as appropriate \cite{o2019physical}. Although assessing patient's performance on rehabilitation exercises is important for therapists to adjust interventions, this assessment relies on therapist's experience \cite{smith2008mispredictions} and infrequently performed due to the limited availability of a therapist. In addition, therapists primarily reply on patient's self-report and do not have any quantitative performance data to understand how well patients follow the prescribed regimens \cite{hendricks2002motor}. Thus, therapists encounter challenges of understanding patient's performance and adjusting intervention. As a first step, this paper primarily focuses on understanding the effect of a machine learning-based decision support system for assessing stroke rehabilitation exercises. 

\subsection{Technological Support for Physical Rehabilitation}\label{sect:related-tech}
To address the limitation of current practices in physical rehabilitation, researchers have explored the feasibility of clinical decision support systems that assist a clinician to obtain insights on patients by monitoring and assessing chronic diseases with computational models \cite{webster2014systematic}.

One approach is called a rule-based model, in which domain experts, clinicians, elicit a set of monitoring rules \cite{siewiorek2012architecture}. For example, Huang explored a tool with therapists to specify repetitions and joint angles for monitoring knee rehabilitation exercises \cite{huang2015exploring}. This rule-based approach provides the modularization and flexibility to develop a monitoring system. However, it is time consuming to determine the right threshold values of rules for an individual's status. Moreover, experts might not be able to articulate their decision making on a complex monitoring task. Alternative approach is a statistical model, which utilizes machine learning with labeled sensor data \cite{siewiorek2012architecture}. This statistical approach utilizes machine learning algorithms to process complex sensor data and automatically extract a meaningful function (e.g. Neural Network model) that can classify the quality of motion \cite{das2011quantitative,lee2019learning}. However, no algorithms can completely replicate therapist's assessment given patient's diverse physical characteristics and functional abilities. Moreover, a statistical approach with complex algorithms cannot explain its prediction to support expert's decision making \cite{gunning2017explainable}, which exacerbate therapist's trust and experience with a decision support system \cite{khairat2018reasons}. 

In this paper, we aim 
to increase the interpretability of a model by feature selection \cite{kim2015mind,biran2017explanation}. Specifically, we apply reinforcement learning \cite{van2016deep,ijcai2019-lee} to identify kinematic salient features for assessment. Utilizing an identified subset of features, we predict the quality of motion and generate user-specific analysis to summarize patient's exercise performance \cite{lee2019intelligent}. 
Our work demonstrates how a tool with predicted assessment and user-specific analysis can affect therapist's decision making on rehabilitation assessment. 


A substantial body of prior work focus on demonstrating the feasibility of collecting objective kinematic variables to quantify the performance of rehabilitation exercises \cite{murphy2011kinematic} and assessing the quality of motion \cite{lee2019learning}. Yet, there is a lack of knowledge and evaluation about therapist's experience on a decision support system for physical rehabilitation monitoring and assessment. Although clinical decision support systems can improve the practices of healthcare \cite{cai2019human}, systems might not be adopted in clinical practices due to lack of user trust and acceptance \cite{devaraj2014barriers,khairat2018reasons}. Specifically, clinical experts might not use a system if it is not properly integrated into workflow and does not provide relevant information \cite{devaraj2014barriers}. This paper contributes to increase knowledge about therapist's needs and experience on an intelligent decision support system for stroke rehabilitation assessment. We conducted a user study with therapists to investigate what types of capabilities therapists want, how therapists use a system, and how user-specific analysis of a system affect therapist's attitude about a system and assessment. 

\section{Stroke Rehabilitation as a Test Domain} \label{sect:study-design}
Stroke is the second leading cause of death and third most common contributor to disability \cite{feigin2017global}. As stroke has increased across the world, we selected stroke rehabilitation as a probe domain. We recruited nine therapists of stroke rehabilitation from five rehabilitation centers (Table \ref{tab:list-tps}) to understand their needs during stroke rehabilitation assessment. Three out of nine therapists specified the design of our study (i.e. exercises and performance components for assessment).  One therapist annotated the dataset to implement a system for evaluation. Two therapists reviewed our implementation before running a user study and other seven therapists participated in the evaluation of our implementation. 

\begin{table}[htp!]
\centering
\caption{The participants of the need finding (needs), of the specification, of the annotation, of reviewing the interface (review), and of the evaluation}
\resizebox{0.8\columnwidth}{!}{%
\begin{tabular}{cccccccc} \toprule
\multirow{2}{*}{ID} & \multicolumn{5}{c}{Studies} & \multirow{2}{*}{\begin{tabular}[c]{@{}c@{}}\# of Years in\\ Stroke Rehab\end{tabular}} & \multirow{2}{*}{Role} \\
 & Needs & Specification & Annotation & Review & Evaluation &  &  \\ \midrule
TP1 & \checkmark & \checkmark & \checkmark & \checkmark &  & 6 & Occupational\\
TP2 & \checkmark & \checkmark &  & \checkmark &  & 4 & Occupational\\
TP3 & \checkmark & \checkmark &  &  & \checkmark & 9 & Occupational\\
TP4 & \checkmark &  &  &  & \checkmark & 4 & Occupational\\
TP5 & \checkmark &  &  &  & \checkmark & 1 & Physio\\
TP6 & \checkmark &  &  &  & \checkmark & 6 & Physio\\
TP7 & \checkmark &  &  &  & \checkmark & 5 & Physio\\
TP8 & \checkmark &  &  &  & \checkmark & 21 & Occupational  \\
TP9 & \checkmark &  &  &  & \checkmark & 11 & Occupational\\ \bottomrule
\end{tabular}%
}
\label{tab:list-tps}
\end{table}

\subsection{Needs during Rehabilitation Assessment}\label{sect:needs}
We interviewed and performed focus group discussion with nine therapists (2 males and 7 females, 29.6 $\pm$ 5.4 years old) with 1 - 20 years of experience in stroke rehabilitation ($\mu=7.44$, $\sigma=5.51$) from five rehabilitation centers to gain knowledge about the current practices and therapists' needs of assessing patient's rehabilitation exercises.
A group of therapists or an individual at each center participated in need finding study for an hour on the same topics: the process of assessment, strategies to cope with an uncertain situation, the current usage of technology, and opportunities for technological support.
In addition, we observed an one hour-long rehabilitation session at one rehabilitation center. Our thematic analysis on need findings with therapists and observation on a rehabilitation session is described as follows:

\subsubsection{Therapist's experience-based and Infrequent Assessment}
Therapists mainly rely on their observation and experience to approximately assess patient's performance on rehabilitation exercises \cite{sanford1993reliability,taub2011wolf} and determine interventions \cite{o2019physical}. When assessing rehabilitation exercises, therapists commented that \textit{``there is no exact single normality for assessment''} (TP 1). Instead, therapists mentioned that they first \textit{``check the functionality of unaffected side and define adequate normality for each patient''} (TP 9). Therapists then internally generate hypothetical correctness of a movement with patient's unaffected side and then \textit{``analyze various aspects of performance: whether a patient can complete an expected movement and any compensated, not coordinated movement exists''} (TP 2). 
During our observation on a rehabilitation session, a therapist first asked a patient to perform a motion multiple times or keep at a certain position for a while for assessment. A therapist then had to keep moving front, back, and side to collect evidences for assessment and expressed a \textit{``difficulty with collecting information on patient's rehabilitation exercise performance''} (TP 3). 

When therapists are unsure, they mentioned that they {record patient's movements to review} and \textit{``re-evaluate more confidently after a session by watching a video multiple times''} (TP 7), or \textit{``discuss with other colleagues''} (TP 8)  on their experience-based assessment. As the process of the assessment is time consuming, therapists only perform infrequently the assessment (e.g. every two or three months).

\subsubsection{Desire for Alternative Perspectives on Assessment with Quantitative Measurements}
All rehabilitation centers that we visited or discussed do not use any technology for managing stroke rehabilitation. When discussing opportunities of technological support to assess rehabilitation exercises, therapists referred the need to gain insights on patient's performance with alternative perspectives on assessment and quantitative kinematic measurements. As mentioned before, therapists have  \textit{``difficulty to detect minor changes over time or discuss with other colleagues''} (TP 2) without quantitative kinematic measurements. In an uncertain situation, a therapist desired to \textit{``validate his/her assessment with alternative assessment from a colleague instead of relying on only my own experience''} (TP 8). Overall, therapists desired a system that can provide \textit{``another perspective of assessment with quantitative measurements''} (TP 6).

Specifically, therapists want to know \textit{``how closely a patient can reach a target motion''} and \textit{``to which extent a patient performs a compensated motion''} (e.g. \textit{``how much a shoulder joint is elevated''}) (TP 3) with quantitative measurements and images of a patient's motion. In addition, therapists desire to understand whether a motion is smooth or not. However, as smoothness has abstract definition, therapists have \textit{``difficulty with assessing smoothness of a motion''} (TP 1). \textit{``Trajectory trends (e.g. showing a graph about how a wrist joint moves during a motion) would be useful to understand smoothness of motion''} (TP 1). 

TP 9 commented that her rehabilitation center attempted to use a system to monitor rehabilitation before, but ended up discard it due its complex and time-consuming process for the usage. For presentation and using a system, TP 9 emphasized that \textit{``a system should be easy to use and present insights quickly with graphics given the limited session time for each patient.''}.  

Based on our need findings with therapists, we have identified the requirements of an intelligent decision support system for stroke rehabilitation in Table \ref{tab:needs}.

\begin{table}[htp]
\caption{Summarized list of needs from therapists and requirements of an intelligent decision support system}
\label{tab:needs}
\resizebox{\textwidth}{!}{%
\begin{tabular}{ll} \toprule
\multicolumn{1}{c}{\textbf{Needs}} & \multicolumn{1}{c}{\textbf{Requirements}} \\ \midrule
N1. Define normality with unaffected motions of a patient & R1. Comparison between unaffected and affected motions \\ \midrule
N2. Validate assessment with another perspective of assessment & R2. Prediction on assessment from a model calibrated with another therapist's assessment \\ \midrule
\begin{tabular}[c]{@{}l@{}}N3. Collect information on patient's performance\\        - N3.1. Detect minor changes\\        - N3.2. Watch a video multiple times\\        - N3.3. Understand smoothness of a motion\end{tabular} & \begin{tabular}[c]{@{}l@{}}R3. Present additional patient-specific analysis\\        - R3.1. Quantitative kinematic measurements \\        - R3.2. Images of a patient's motion\\        - R3.3. Joint trajectory trends\end{tabular} \\ \midrule
N4. Simple and intuitive presentation & R4. Avoid overwhelming therapists and utilize graphics to present insights quickly \\ \bottomrule
\end{tabular}%
}
\end{table}

\subsection{Specifications}
After having iterative discussion with three therapists (with $\mu = 6.49$, $\sigma=2.05$ years of experience in stroke rehabilitation), we specified exercises and performance components of assessment to probe how therapists utilize an intelligent decision support system to assess patient's rehabilitation exercises. 

\subsubsection{Three Task-Oriented Upper Limb Exercises} 
\begin{figure}[th]
\centering
\begin{subfigure}{.32\columnwidth}
\centering
  \includegraphics[width=0.8\columnwidth]{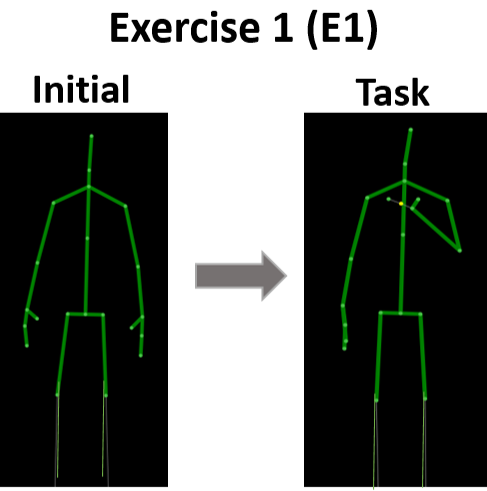}
  \caption{}
  \label{fig:exercise1}
\end{subfigure} 
\begin{subfigure}{.32\columnwidth}
  \centering
  \includegraphics[width=0.8\columnwidth]{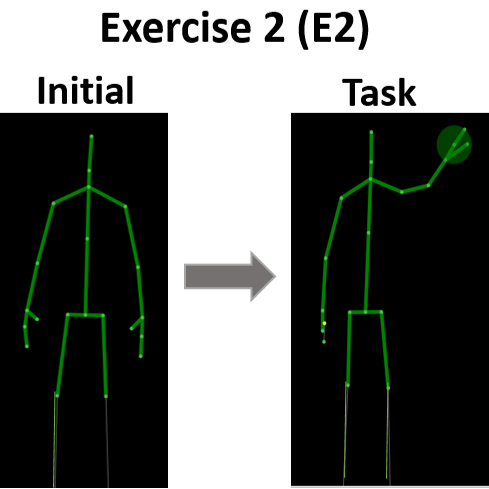}
  \caption{}
  \label{fig:exercise2}
\end{subfigure}
\begin{subfigure}{.32\columnwidth}
  \centering
  \includegraphics[width=0.8\columnwidth]{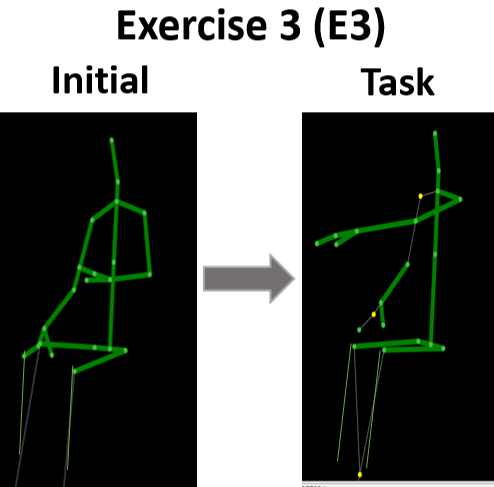}
  \caption{}
  \label{fig:exercise3}
\end{subfigure}
\caption{(a) Exercise 1 (E1): \textit{`Bring a Cup to the Mouth'} (b) Exercise 2 (E2): \textit{`Switch a Light On'} (c) Exercise 3 (E3): \textit{`Move a Cane Forward'}}~\label{fig:three-exercises}
\end{figure}

This paper utilizes three upper-limb stroke rehabilitation exercises (Figure \ref{fig:three-exercises}), recommended by therapists \cite{lee2018technology}. In Figure \ref{fig:three-exercises}, the \textit{`Initial'} indicates the initial position of an exercise and the \textit{`Target'} describes the desired end position of an exercise. 

For Exercise 1, a subject has to raise his/her wrist to the mouth as if drinking water. For Exercise 2, a subject has to pretend touching a light switch on the wall. Exercise 3 is to practice the usage of a cane while extending elbow in the seated position. These exercises are selected due to their correspondence with major motion patterns: elbow flexion for Exercise 1, shoulder flexion for Exercise 2, elbow extension for Exercise 3. 

\subsubsection{Performance Components}
After reviewing popular stroke assessment tools (i.e. Fugl Meyer Assessment \cite{sanford1993reliability} and Wolf Motor Function Test \cite{taub2011wolf}) and having iterative discussion with therapists, we identified three common performance components and their scoring guidelines: \textit{`Range of Motion (ROM)'}, \textit{`Smoothness'}, and  \textit{`Compensation'} (Table \ref{tab:score-guidelines}). The \textit{`ROM'} component describes the amount of a joint movement to achieve a task-oriented exercise. The \textit{`Smoothness'} component indicates the degree of a trembling and irregular movement of joints while performing an exercise. The \textit{`Compensation'} component checks whether compensated movements are used to achieve a target movement. For instance, a patient might elevate his/her shoulder to raise the affected hand as shown in Figure \ref{fig:sample-compensation-abnormal-p11} and \ref{fig:sample-compensation-abnormal-p14}.

\begin{table}[htp!]
\centering
\resizebox{0.7\columnwidth}{!}{%
\begin{tabular}{ccl} \toprule
\textbf{\begin{tabular}[c]{@{}c@{}}Performance \\ Components\end{tabular}} & \textbf{Score} & \multicolumn{1}{c}{\textbf{Guidelines}} \\  \midrule
\multirow{3}{*}{\begin{tabular}[c]{@{}c@{}}Range of Movement \\ (ROM) 
\end{tabular}
} & 0 & Does not or barely involve any movement \\
 & 1 & Less than half way aligned with an \textit{`Target'} position
 \\
 & 2 & Movement achieves an \textit{`Target'} position \\ \midrule
\multirow{3}{*}{Smoothness} & 0 & Excessive tremor or not smooth coordination \\
 & 1 & Movement influenced by tremor \\
 & 2 & Smoothly coordinated movement \\ \midrule
\multirow{3}{*}{Compensation} & 0 & Noticeable compensation in more than two joints \\
 & 1 & Noticeable compensation in a joint \\
 & 2 & Does not involve any compensations \\ \bottomrule
\end{tabular}%
}
\caption{Guidelines to Assess Stroke Rehabilitation Exercises}
\label{tab:score-guidelines}
\end{table}

\subsubsection{Kinematic Features}
We utilize an exercise dataset that is composed of sequential joint coordinates of motions and extract various kinematic features. To represent the \textit{`ROM'} component, we extract joint angles (e.g. elbow flexion, shoulder flexion, elbow extension), normalized relative trajectory (i.e. Euclidean distance between two joints - head and wrist, head and elbow), and normalized trajectory distance (i.e. absolute distance between two joints - head and wrist, shoulder and wrist) in x, y, z coordinates. 

For the \textit{`Smoothness'} component, we compute the speed, acceleration, and jerk on wrist and elbow joints. Moreover, normalized speed and acceleration, and Mean Arrest Period Ratio (the portion of the frames when speed exceeds $10\%$ of the maximum speed) are also included based on the prior work \cite{rohrer2002movement}. 

For the \textit{`Compensation'} component, we compute joint angles (i.e. the elevated angle of a shoulder, the tilted angle of spine, and shoulder abduction) and normalized trajectories (the distance between joint positions of head, spine, shoulder joints in x, y, z axis from the initial to the current frames) to distinguish a compensated movement. 

Before extracting features, we apply a moving average filter with the window size of five frames to reduce noise of acquiring joint positions from a Kinect sensor similar to \cite{stone2011evaluation}. 
For each exercise motion, we compute a feature matrix ($\textbf{F} \in R^{t \times d}$) with $t$ frame and $d$ features and statistics (i.e. max, min, range, average, and standard deviation) over all frames of the exercise to summarize a motion. 

\section{Intelligent Decision Support System for Stroke Rehabilitation Assessment}\label{sect:method}
Based on identified therapists' needs, we designed and implemented an intelligent decision support system (Figure \ref{fig:overall-arch}) that can identify salient features for assessment using reinforcement learning to predict the quality of motion and generate user-specific analysis that includes feature analysis, images of salient frames, and trajectory trends (Figure \ref{fig:web-interface}). This system enables therapists to review alternative perspectives of patient's performance with quantitative user-specific analysis for assessment.

\subsection{Prediction Model}
The Prediction Model (PM) applies a supervised learning algorithm to predict the quality of motion on each performance component. We explore various traditional supervised learning algorithms: Decision Trees (DTs), Linear Regression (LR), Support Vector Machine (SVM) using the \textit{`Scikit-learn'} \cite{pedregosa2011scikit} library and Neural Networks (NNs) using \textit{`PyTorch'} \cite{paszke2017automatic} library. 

For DTs, we implement Classification and Regression Trees (CART) to build prune trees. For LR models, we apply $L1, L2$ regularization or linear combination of $L1$ and $L2$ (ElasticNet with $0.5$ ratio) to avoid overfitting. For SVMs, we apply either linear or  Radial Basis Function (RBF) kernels with penalty parameter, $C = 1.0$. For NNs, we grid-search various architectures (i.e. one to three layers with $32, 64, 128, 256, 512$ hidden units) and an adaptive learning rate with different initial learning rates (i.e. $0.0001, 0.005, 0.001, 0.01, 0.1$). We apply \textit{`ReLu'} activation functions and \textit{`AdamOptimizer'} and train a model until the tolerance of optimization is $0.0001$ or the maximum $200$ iterations.

\subsection{Feature Selection using Reinforcement Learning}
Kinematic variables analysis is an important way for therapists to quantitatively understand patient's performance \cite{wu2000kinematic}. Yet, simply presenting all variables can overwhelm therapists and limit therapist's ability to gain insights on patient's performance. Given the limited availability to administrate multiple patients, therapists want to minimize the amount of time on analyzing kinematic variables while accurately diagnosing patient's status. Thus, we aim at automatically identifying salient features of assessment with machine learning. 

The classical approaches of feature selection (e.g. filter, wrapper, embedded methods) \cite{tang2014feature} find a fixed feature set to the entire dataset, which applies globally to all patients. Instead, this paper utilizes a Markov Decision Process (MDP) to select a feature set for each patient's motions. As each patient has different physical and functional status (Figure \ref{fig:sample-compensation}), we hypothesize that feature selection with MDP can be beneficial over classical feature selection approaches for personalized rehabilitation assessment. 


\subsubsection{Problem Definition}
We formulate this problem of feature selection as a Markov Decision Process (MDP), where each episode is to classify an instance and the environment is the power set of the feature space. An agent sequentially determines whether to query additional feature or classify a sample while receiving a negative reward for recruiting a feature or mis-classification. 
To solve this problem, we apply Deep Q-network with Double Q-learning \cite{mnih2015human,van2016deep} with the same architectures of Neural Network for the Prediction Model (PM) (Table \ref{tab:params}) using \textit{`PyTorch'} libraries \cite{paszke2017automatic}.



We mathematically describe the Markov Decision Process (MDP) with similar notations of \cite{dulac2011datum,janisch2019classification} as follows:

Let $(x, y) \in \mathcal D = \mathcal{X}\times \mathcal{Y}$ be a sample from a dataset, where x is a feature vector and y is the class label.  Let $\mathcal{F}$ be the set of identified features and the function $c:\mathcal{F}\to$ $\mathbb{R}^{\leq0}$ be the cost of adding a feature in $\mathcal{F}$. 

\begin{itemize}
    \item \textbf{State Space}: $\mathcal{S}$, let state $s = (x,y,\mathcal{F}) \in \mathcal{S} = \mathcal{X}\times \mathcal{Y} \times \mathcal{P(F)}$, and the observed state without the label be \\$s'= \{(x_{i},f_i)\mid \forall$ features $f_i \in \mathcal{F} \}$.
    \item \textbf{Action Space}: Let $\mathcal{A}$ denote the action set. The agent takes either action $a_{c}$ or $a_{f_i} \in \mathcal{A}$, where $a_{c}$ classifies the instance and $a_{f_i}$ queries feature $f_i$. 
    \item \textbf{Reward}: Let the reward function be defined as 
$r(s,a) = r((x,y,\mathcal{F}),a)= $
$\begin{cases} c(f_i) & \text{if } a_{f_i}\\ -1 & \text{if } a_{c} \text{ and } a_{c} \neq y \\0 & \text{if } a_{c} \text{ and } a_{c}=y\end{cases}$\\

We apply an uniform cost over features: $\forall f_i, \ c(f_i)=- \lambda$, where $\lambda$ = 0.01. 

\item \textbf{Transition}: Let the transition function be $p(s,a)= p((x,y,\mathcal{F}), a) = \begin{cases} (x,y,\mathcal{F}\cup f_i) & \text{if } a_{f_i}\\ T & \text{if } a_{c} \end{cases}$, \\where T is the terminal state after outputting the classification and revealing the true label.
\end{itemize}

\subsection{Visualization Interface}\label{sect:method-intf}
Based on the therapists' needs (Section \ref{sect:needs})  and few guidelines of Human Artificial Intelligence (AI) interaction \cite{kulesza2015principles,amershi2019guidelines}, we implement the web-based visualization interface that presents a predicted performance score and user-specific analysis that contains feature analysis, images of salient frames, and trajectory trends (Figure \ref{fig:web-interface}) to validate and support therapist's assessment. Feature analysis shows quantitative difference between unaffected and affected sides using identified salient features (Figure \ref{fig:web-feat}). An image of salient frames shows patient's motion at salient frames, in which salient features are occurred (Figure \ref{fig:web-img}). Trajectory analysis describes trends and duration of a joint trajectory during a motion (Figure \ref{fig:web-traj}). 
The interface has the tab menus to present videos and each analysis respectively. We implement the javascript functions to count the video events (e.g. \textit{`Play'}, \textit{`Pause'}) and measure the amount of time that a user spends on each page during assessment. 



 
As therapists want another perspectives on assessment to validate their own assessment ($N2$ in the Table \ref{tab:needs}), this interface presents the predicted assessment, scores on performance components. 
When presenting this predicted performance score, the performance of predictions is also included to \textit{``make clear how well the system can do''}  \cite{amershi2019guidelines}. In addition, this interface presents user-specific analysis that is considered \textit{``contextually relevant information''} \cite{amershi2019guidelines} on patient's exercise performance ($N3$ in the Table \ref{tab:needs}) from therapists. Specifically, user-specific analysis of our interface includes the presentation of feature analysis, images of salient frames, and trajectory trends that are identified during the needs finding study. 
User-specific analysis with identified kinematic features is referred as explanations of predicted assessment through out this paper as described in the Section \ref{sect:related-tech}.

For simple and intuitive presentation ($N4$ in Table \ref{tab:needs}) on quantitative measurements of identified salient features, this interface utilizes a radar chart to effectively present multivariate data. 
To \textit{``avoid overwhelming''} \cite{kulesza2015principles} therapists, this interface limits to include only three salient features with highest information gain. 
Utilizing selected salient features (e.g. the maximum target position, maximum elbow flexion), we identify frames in which these salient features are occurred to present images. In addition, as observing sequential patterns of kinematic variables provides another useful perspective in some cases (e.g. the assessment of the \textit{`Smoothness'} performance component), this interface shows trajectories of three major joints (e.g. shoulder, elbow, and wrist) for upper-limb exercises. As therapists utilize patient's unaffected motion as normality to assess patient's performance ($N1$ in Table \ref{tab:needs}), this interface follows this current practice, \textit{``social norms''} \cite{amershi2019guidelines} and includes the comparison between the affected and unaffected side to present salient features and trajectory trends. 

\section{Experiment for System Implementation}
\subsection{Data Collection}
We recruited 15 stroke patients and 11 healthy subjects to collect the dataset of three upper limb exercises using a Kinect v2 sensor (Microsoft, Redmond, USA). The data collection program is implemented in C\# using Kinect SDK and operated on a PC with 8GB RAM and i5-4590 3.3GHz 4 Cores CPU. This program records the 3D trajectory of joints and video frames at 30 Hz. The sensor was located at a height of 0.72m above the floor and 2.5m away from a subject. The starting and ending frames of exercise movements were manually annotated during the data collection.

Before participating in the data collection, all subjects signed the consent form. Fifteen post-stroke patients (13 males and 2 females) participated in two sessions for data collection: During the first session, a therapist evaluated post-stroke patient's functional ability using the a clinically validated tool, Fugl Meyer Assessment (FMA) (the maximum score on 66 points) \cite{sanford1993reliability}. Fifteen stroke survivors have diverse functional abilities from mild to severe impairment (37 $\pm$ 21 Fugl Meyer Scores). During the second session, a stroke survivor performed 10 repetitions of each exercise with both affected and unaffected sides. Eleven healthy subjects (10 males and 1 female) performed 15 repetitions with their dominant arms for each exercise. 

\textbf{\textit{`Training Data'}} (Figure \ref{fig:overall-arch}) is composed of 165 unaffected motions from 11 healthy subjects and 150 affected motions from 15 stroke survivors to train the Prediction Model (PM). 

\textbf{\textit{`User Data'}} (Figure \ref{fig:overall-arch}) includes unaffected and affected motions of a testing stroke subject to generate user-specific analysis of the visualization interface (Figure \ref{fig:web-interface}).

\subsection{Annotations and Design Review on Interface}
For implementation, we utilize the annotation of therapist 1 (TP 1), who had more interactions with recruited stroke patients by supporting the recruitment and evaluation on their functional ability with Fugl Meyer Assessment. Therapist 1 (TP1) watched the recorded videos of patient's movements (Figure \ref{fig:web-video}) and annotated exercise motion dataset using the scoring guideline (Table \ref{tab:score-guidelines}) without reviewing analysis of our system (Figure \ref{fig:web-traj}, \ref{fig:web-img}, \ref{fig:web-traj}).

After implementing the system and interface, therapist 1 and 2 (TP 1 and 2) reviewed the web interface to detect any problems and improve its usability. According to the review, TP 1 and 2 had problems with understanding the name of features in feature analysis. Thus, we reviewed the names of features with TP 1 and 2 and converted them into clinically relevant terminologies. For this conversion of feature names, we presented all feature names and described what each feature measures to TP 1 and 2. They spoke aloud how they would describe each feature. For instance, \textit{`Normalized trajectory distance of spine x'} is converted to \textit{`Leaning trunk to the side'}.

\section{Real-World User Study}
We performed a user study to investigate how the information of an intelligent decision support system (e.g. predicted performance scores with feature analysis, images of salient frames, trajectory trends) affect therapist's rehabilitation assessment. 
For the user study, we compared the experiences of therapists using our proposed interface (Figure \ref{fig:web-interface}) to two baseline interfaces: \textit{`Traditional'} interface that presents only videos for assessment and \textit{`Predicted Scores'} interface that presents videos with predicted scores without any user-specific analysis. 
Specifically, we aim to address the following questions:

\begin{itemize}
    \item RQ 1: How do predicted assessment and user-specific analysis of our tool affect the utility of information, workload, and trust, compared to two baseline interfaces (one with only videos and the other with videos and only predicted assessment)? Do predicted assessment and user-specific analysis of our tool support more consistent assessment?
    \item RQ 2: How do therapists utilize each user-specific analysis for assessment?
\end{itemize}

\subsection{Metrics}
We evaluated three interfaces with respect to the following metrics: 1) subjective feedback on questionnaires, 2) logs of the web-based visualization interface, 3) agreement level of therapists' evaluation (F1-scores).

\subsubsection{Subjective Feedback on Questionnaires}
We utilize the following questionnaires \cite{cai2019human} to collect therapist's subjective feedback on interfaces. All questionnaires were rated on a 7-point scale.

\begin{itemize}
    \item Usefulness: \textit{``[Tool - Condition X] is useful to understand and assess patient's performance''}
    \item Richness: \textit{``[Tool - Condition X] generates new insights on patient's performance''}
    \item Trust: \textit{``I can trust information from [Tool - Condition X]''}    
    \item Workload: participants answered the \textit{``efforts''} and \textit{``workload''} dimensions of the NASA-TLX \cite{hart1988development}
    \item Usage Intention: \textit{``I would use [Tool - Condition X] to understand and assess patient's performance''}
    \item Preference between two interfaces: participants rated on a 7-point scale ranging from 1 (totally Condition X), 2 (much more Condition X than Y), 3 (slightly more Condition X than Y), 4 (neutral), ..., 7 (totally Condition Y).
\end{itemize}

The preference is asked pairwise on three conditions/interfaces: Condition 1 (\textit{`Traditional'} interface), Condition 2 (\textit{`Predicted Scores'} interface), and Condition 3 (\textit{`Proposed'} interface). 

\subsubsection{Logs of the Web Interface}
Our web interfaces record a log file that counts the number of video events (e.g. \textit{`Play'}, \textit{`Pause'}) and measures the amount of time that a participate spends on each page/resource during assessment. 

\subsubsection{Agreement Level of Therapists' Evaluation}
Participants generate assessment of patient's exercise performance using the interfaces. To understand whether our proposed tool with user-specific analysis supports more consistent evaluation, we compute the level of agreement of therapists' evaluation (F1-score) for each interface. 



\subsection{Method}
Seven therapists (with $\mu = 8.14$, $\sigma = 6.05$ years of experience in stroke rehabilitation) from four rehabilitation centers participated in the user study on the evaluation (Table \ref{tab:list-tps}). Note that we excluded two therapists (TP 1 and 2), who generated annotation to implement our system and reviewed the design of the interface. After signing an informed consent (Institutional Review Board approved), each participant was instructed on the procedure of the study and three interfaces using dummy data (30 minutes). Then, a participant is assigned the task of assessing 45 videos (around one minute per video, in which a patient performs a rehabilitation exercise) using three interfaces (1.5 hours total) and followed by post-study questionnaires and interview (30 minutes). 

Each interface is assigned a sub-task of assessing 15 videos (five patients performing three exercises). Therapists 1, who evaluated functional ability of 15 patients, divided 15 patients into three sub-groups, in which patients of each subgroup have similar functional ability. Thus, the sub-task of each interface is counterbalanced. The order of the three conditions/interfaces and assignment of a sub-task are randomized. After completing a sub-task on each interface, therapists responded the questionnaires. After finishing all sub-tasks, therapists answered the preference questionnaires and post-interview is conducted to understand therapists' perspectives on the effectiveness of the proposed, intelligent decision support system and the opportunities to utilize this system in the practice.



\section{System Implementation Results}

To evaluate our implementation of the Prediction Model (PM), we apply Leave-One-Subject-Out (LOSO) cross validation on post-stroke patients, which trains data from all subjects except one post-stroke survivor and test with data from the left-out post-stroke survivor. Table \ref{tab:results-avg} summarizes average F1-scores of models of three exercises. 

\begin{table}[htp]
\centering
\resizebox{0.8\columnwidth}{!}{%
\begin{tabular}{ccccc} \toprule
 & Exercise 1 (E1) & Exercise 2 (E2) & Exercise 3 (E3) & Overall \\ \toprule
PM - DT & 0.6901  $\pm$ 0.0405 & 0.7645  $\pm$ 0.0867 & 0.6488 $\pm$ 0.0412 & 0.7011 $\pm$ 0.0769\\ \midrule
PM - LR & 0.7246 $\pm$ 0.0593 & 0.6430 $\pm$ 0.0982 & 0.7267 $\pm$ 0.0391 & 0.6981 $\pm$ 0.0801\\ \midrule
PM - SVM & 0.7232 $\pm$ 0.0364 & 0.6971  $\pm$ 0.0891 & 0.7410 $\pm$ 0.0052 & 0.7204 $\pm$ 0.0585\\ \midrule
PM - NN & \textbf{0.8806  $\pm$ 0.0502} & \textbf{0.8090  $\pm$ 0.0671} & \textbf{0.8115 $\pm$ 0.0436} & \textbf{0.8337 $\pm$ 0.0638}\\ \bottomrule
\end{tabular} 
}
\caption{Performance of Prediction Model (PM), agreement with therapist 1's evaluation (F1-scores) using Decision Tree (DT), Linear Regression (LR), Support Vector Machine (SVM), and Neural Network (NN)}
\label{tab:results-avg}
\end{table}

The Prediction Model (PM) using Neural Networks (NNs) achieves decent agreement level with Therapist 1's evaluation: 0.8337 average F1-scores over three exercises. In addition, the PM with NNs outperforms the PM with other algorithms: Decision Trees (0.7011 average F1-scores), Linear Regression (0.6981 average F1-scores), Support Vector Machine (0.7204 average F1-scores). The parameters of NNs (i.e. hidden layers/units and learning rate) that achieve the best F1-score on the classification are summarized in the Table \ref{tab:params} in the Appendix. 

\begin{figure} [htp]
\centering
  \includegraphics[width=0.7\columnwidth]{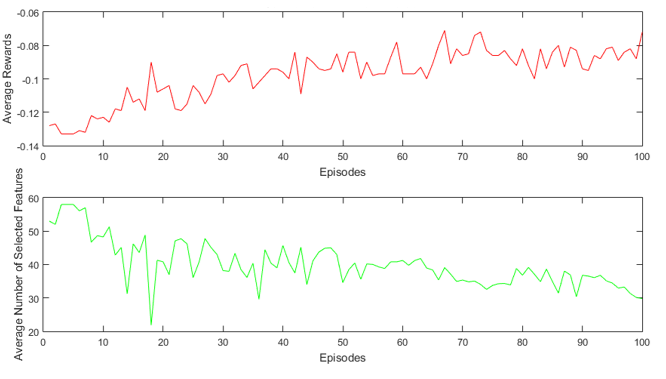}
  \caption{The average rewards and the average number of selected features while training an agent for feature selection}~\label{figures:drl-validation}
\end{figure}

We found that {our system can identify salient features of assessment for individual patient's motions}. Utilizing the Neural Network architectures of the PM (Table \ref{tab:params}), we train an agent that sequentially decides whether another feature is necessary to assess the quality of motion. To validate the feasibility of our implementation, we plot the average rewards and the average number of selected features during training an agent. Figure \ref{figures:drl-validation} demonstrates that an agent can identify the salient subset of features while reducing the number of selected features and improving average rewards (i.e. the correct assessment of exercise motions). In addition, compared to a model with Recursive Feature Elimination (RFE) method \cite{guyon2002gene}, one of classical feature selection methods, our approach has 0.11 higher average F1-score ($p < 0.01$ using a paired t-test over 3 exercises and 3 components) and is expected to be more beneficial to generate patient-specific analysis for therapists. 

\section{User Study Results}

\subsection{Subjective Feedback on Questionnaires}
\begin{figure} [ht!]
\centering
  \includegraphics[width=0.8\columnwidth]{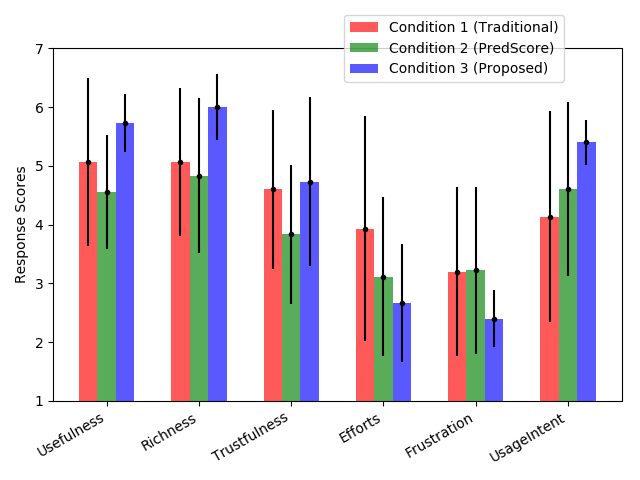}
  \caption{Results of questionnaires on three interfaces. The proposed interface, Condition 3 is more useful, richer, more trustful while reducing efforts and frustration on assessment task. It is more likely to be used in the clinical practices than other baseline interfaces }~\label{fig:results-questionaires}
\end{figure}

Figure \ref{fig:results-questionaires} summarizes the responses of questionnaires from therapists on three interfaces: Condition 1 (\textit{`Traditional'} interface), Condition 2 (\textit{`Predicted Scores'} interface), and Condition 3 (\textit{`Proposed'} interface). 


Our proposed interface, Condition 3 achieves \textbf{higher usefulness} ($\mu = 5.73$) than the others (Condition 1: $\mu = 5.06$, $p = 0.15$ and Condition 2: $\mu = 4.55$, $p < 0.05$) and \textbf{higher richness} ($\mu = 6.00$) than the others (Condition 1: $\mu = 5.06$, $p = 0.06$ and Condition 2: $\mu = 4.83$, $p = 0.06$) as additional explanations of Condition 3 were considered \textit{``useful to understand patient's condition''} for therapists. In addition, participated therapists expressed  \textbf{higher trust} on Condition 3 ($\mu = 4.73$) than the others (Condition 1: $\mu = 4.60$, $p = 0.42$ and Condition 2: $\mu = 3.833$, $p = 0.18$). Although participated therapists identified that some \textit{``predicted scores are not trustful''}, analysis of the proposed interface complements to \textit{``understand why such predicted scores is generated''}. 

Participated therapists experienced \textbf{lower efforts} ($\mu = 2.66$) using the proposed, Condition 3 than the others (Condition 1: $\mu = 3.93, p = 0.08$ and Condition 2: $\mu = 3.11$, $p = 0.27$) and \textbf{lower frustration} ($\mu = 2.40$) using the proposed, Condition 3 than the others (Condition 1: $\mu = 3.20$, $p = 0.12$ and Condition 2: $\mu = 3.22$, $p = 0.07$). Participated therapists described that user-specific analysis of the proposed interface (feature analysis, images of salient frames, and trajectory trends) reduce the effort and frustration to \textit{``search evidences in videos''}.

In addition, the proposed, Condition 3 interface has \textbf{higher usage intent} ($\mu = 5.4$) than the others (Condition 1: $\mu = 4.13$, $p < 0.05$ and Condition 2: $\mu = 4.61$, $p = 0.10$) and therapists mostly \textbf{prefer the proposed, Condition 3} interface to two baseline interfaces (Condition 1 and 2): for preference questionnaire between Condition 1 and 3, 2 out of 7 therapists \textit{`Totally'} prefer Condition 3, 4 out of 7 therapists \textit{`Much more preferred'} Condition 3 than Condition 1, and 1 out of 7 therapists \textit{`Much more preferred'} Condition 1 than Condition 3. For preference questionnaire between Condition 2 and 3, 4 out of 7 therapists \textit{`Totally'} prefer Condition 3, 2 out of 7 therapists \textit{`Much more preferred'} Condition 3 than Condition 2, and 1 out of 7 therapists \textit{`Slightly more preferred'} Condition 2 than Condition 3. Although one therapist still considered the usefulness of the proposed interface, this therapist preferred the interface without predicted assessment and user-specific analysis. 
Participated therapists commented that Condition 3 with predicted assessment and additional user-specific analysis \textit{`is very interesting'} and \textit{`gives me insights to assess patient's performance'}.

In summary, Condition 3 with predicted assessment and patient-specific analysis achieved positive responses on all aspects: our proposed interface provides more useful and richer information to understand patient's performance and has higher trust in the system, reduces therapist's efforts and frustration to find evidences for assessment, and more likely to be used in clinical practices. However, score differences with the baseline interface are not statistically significant except for \textbf{usefulness} and \textbf{usage intent} aspects. 

\subsection{Logs of the Web Interface}
Table \ref{tab:results-logs-stats} describes the measurements of logs (i.e. average number of video events and time on video/analysis per assessment) from three interfaces. Our proposed interface, Condition 3 has \textbf{significantly lower, average number of video events} ($\mu = 4.25$) than the others (Condition 1: $\mu = 6.45$ and Condition 2: $\mu = 10.16$) ($p < 0.05$). This lower number of video events on Condition 3 indicates that Condition 3 leads to lower number of video playbacks than others.

Condition 3 with additional user-specific analysis has {longer average time on assessment} ($\mu = 95.24$ seconds) than Condition 1 ($\mu = 79.43$) and {lower average time on assessment} ($\mu = 95.24$ seconds) than Condition 2 ($\mu = 97.69$ seconds). 
However, when we analyze average time on each information, specifically videos, Condition 3 shows \textbf{significantly lower average time on videos} ($\mu = 46.53$ seconds) than the others (Condition 1: $\mu = 79.43$ seconds and Condition 2: $\mu = 81.97$ seconds) $(p < 0.05)$.


\begin{table}[h]
\centering
\caption{Measurements of the logs on three interfaces}
\resizebox{0.7\columnwidth}{!}{%
\begin{tabular}{ccccl}
\toprule
Interface & \begin{tabular}[c]{@{}c@{}}Average \# of\\ Video Events\end{tabular} & \begin{tabular}[c]{@{}c@{}}Avg. Time on\\ Assessment\\ (seconds)\end{tabular} & \begin{tabular}[c]{@{}c@{}}Avg. Time on\\ Videos\\ (seconds)\end{tabular} & \multicolumn{1}{c}{\begin{tabular}[c]{@{}c@{}}Avg. Time\\ on Information\\ (seconds)\end{tabular}} \\ \midrule
\textit{\begin{tabular}[c]{@{}c@{}}Condition 1\\ ('Traditional')\end{tabular}} & 6.45 & \textbf{79.43} & 79.43 & \multicolumn{1}{c}{N/A} \\ \midrule
\textit{\begin{tabular}[c]{@{}c@{}}Condition 2\\ ('Videos + \\ PredictedScore ')\end{tabular}} & 10.16 & 97.69 & 81.97 & PredictedScore: 15.72 \\ \midrule
\textit{\begin{tabular}[c]{@{}c@{}}Condition 3\\ ('Videos \\ + PredictedScore \\ + Analysis')\end{tabular}} & \textbf{4.25} & 95.24 & \textbf{46.53} & \begin{tabular}[c]{@{}l@{}}Feature      : 22.70\\ SalientFrames: 14.39\\ Trajectory: 17.51\end{tabular} \\ \midrule
\end{tabular}%
}
\label{tab:results-logs-stats}
\end{table}

\subsection{Agreement Level of Therapists' Evaluation}
\begin{figure} [h!]
\centering
  \includegraphics[width=0.7\columnwidth]{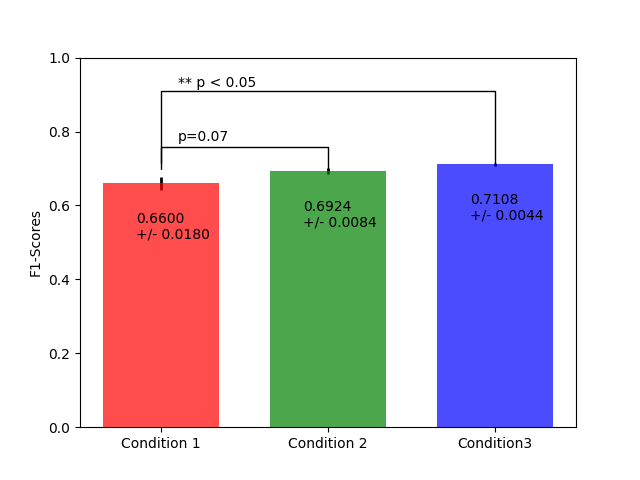}
    \caption{The agreement level (F1-scores) of therapists' evaluation using three interfaces: although both Condition 2 and 3 have higher agreement level than Condition 1, the difference is only significant on Condition 3 with user-specific analysis}~\label{fig:results-agreement}
\end{figure}

Figure \ref{fig:results-agreement} shows the agreement level of therapists' evaluation on three interfaces. Our proposed interface, Condition 3 with predicted assessment and user-specific analysis (i.e. feature analysis, salient frames, and trajectory) achieves \textbf{higher agreement} on participated therapists' evaluation ($\mu = 0.7138$ F1-scores) than the others (Condition 1: $\mu = 0.66$ F1-score ($p < 0.05$) and Condition 2: $\mu = 0.6924$) F1-score. Although both Condition 2 and 3 achieve higher agreement level than Condition 1, the difference between Condition 1 and 2 is not statistically significant ($p = 0.07$), but the difference between Condition 1 and 3 is statistically significant ($p < 0.05$). Thus, this indicates the positive effect of including user-specific analysis on Condition 3 to improve the agreement level of therapists' evaluation. 

\subsection{Post-study Interviews}\label{sect:post-interviews}
After completing the study, we collected general feedback on our proposed interface (Condition 3) from therapists. Specifically, we asked therapist's opinions and usage on the interface and the possibility of accepting it in the current practices. 

Overall, therapists consider the proposed interface as \textit{``a good platform''} (TP 5) for rehabilitation assessment. User-specific analysis of the interface (e.g. feature analysis, images of salient frames, and trajectory trends) \textit{``brings more interesting, new aspects of a patient''} (TP 3) and enables therapists to \textit{``understand why the predicted assessment is suggested''} (TP 9). Specifically, therapists found that {feature analysis} (Figure \ref{fig:web-feat}) is \textit{``easy and intuitive''} (TP 9) to \textit{``quickly observe the quantitative difference between unaffected and affected sides''} (TP 6) for assessment. 
{Images of salient frames} (Figure \ref{fig:web-img}) \textit{``was helpful to validate feature analysis''} (TP 9). {Trajectory trends} (Figure \ref{fig:web-traj}) \textit{``was useful to review the overall trends''} (TP 8) and \textit{``the duration of a motion''} (TP 9), which were \textit{``helpful to assess the smoothness of a motion''} (TP 4). 

Therapists described two different patterns of using the interface for the assessment. One pattern is to first \textit{``review the feature analysis to get the overview of quantitative difference between unaffected and affected side''} (TP 7) and \textit{``validate quantitative feature analysis with images of salient frames''} and \textit{``trajectory trends''} (TP 9). Some therapists preferred to get the initial insight of assessment from feature analysis, because \textit{`` graphics on quantitative difference between unaffected and affected sides are useful and fast to get insights and validate my assessment''} (TP 7). The other strategy is to first \textit{``observe the trajectory trends to understand the overview of a motion''} (TP 8) and \textit{``review the detailed, quantitative feature analysis''} and \textit{``images of salient frames''} (TP 8). Others reviewed trajectory analysis first, because it \textit{``provides various insights (e.g. duration, amplitude, and tremor) together''} to improve and validate therapist's hypothetical assessment. 

After reviewing predicted assessment and user-specific analysis, therapists were able to determine whether a system makes a mistake or not and understand the capabilities of a system. Even if the predicted scores of an interface sometimes mismatch with therapist's assessment, therapists consider \textit{``the proposed interface is trustful''} (TP 9) in a way that \textit{``I can review patient-specific analysis to understand whether a system fails to predict correctly or I make a mistake''} (TP 4). For example, TP 9 commented that \textit{``the prediction on range of motion (ROM) seemed to be aligned most of time with my hypothetical assessment and insights from user-specific analysis''}. In contrast, TP 9 mentioned that predictions of compensation do not sometimes perform well, because the system \textit{``does not provide a prediction that is aligned with mine and include leaning trunk to the side''} feature to predict compensation of a patient, who \textit{``compensates trunk to the side''}. 
Overall, therapists developed a mental model that they would trust/rely more on the prediction of ROM and less on the prediction on Smoothness and Compensation.
Thus, user-specific analysis of our interface assisted therapists to understand the capabilities of a system to predict performance components and develop the different levels of trust on predictions of each performance components from a system.


In addition, therapists considered the user-specific analysis of the proposed interface, Condition 3 \textit{``reduces their efforts and frustration on the assessment''} (TP 6). \textit{``When only video is presented''} (Condition 1 - \textit{`Traditional'}), therapists have  \textit{``difficulty to consider different perspectives on assessment at the same time} (TP 9). So, therapists \textit{``had to replay a video multiple times''} (TP 3). In contrast, the proposed interface provides the insights on patient's performance (i.e. predicted scores and user-specific analysis), which reduce therapist's efforts and frustration to repeatedly watch videos and search evidences for assessment. 
Therapists considered that user-specific analysis on each performance component is useful to reduce complexity of assessment. \textit{``It is complex and challenging process to simultaneously reviewing multiple aspects of assessment while watching a video. In contrast, user-specific analysis on each component from the interface simplified my assessment process''} (TP 9). In addition, TP 9 commented that after getting used to user-specific analysis of our interface, \textit{``I started reducing replay a video to search clues for assessment''} and \textit{``relied more on analysis of the interface''}, because the interface \textit{``quickly presents various quantitative measurements for assessment, which I have to speculate while watching a video''}. 

As the proposed interface is \textit{``easy to use''} and \textit{``quickly summarizes quantitative data with graphics to provide insights of patient's performance''} (TP 9), therapists are positive to accept the interface in their practices. Therapists commented that currently they \textit{``do not have much quantitative data to analyze and discuss with patients''} (TP 7). Therapists described that predicted assessment and user-specific analysis with quantitative data from our interface could facilitate \textit{``understanding on patient's performance and communication it with patients''} (TP 4). In addition, some therapists consider the interface might be \textit{``helpful to motivate patient's participation in rehabilitation program''} (TP 9) by tracking and presenting patient's progress with quantitative data. 


\section{Discussion}
In this section, we synthesize our findings, discuss design recommendations to create a decision support system for rehabilitation assessment: the importance of 1) presenting appropriate and salient data and 2) deriving an adaptive system for a personalized human and machine collaborative decision support system, and describe the limitations of our study.


\subsection{Design Recommendations}
\subsubsection{Presenting Appropriate and Salient Data to Understand Capabilities of a System}
Clinical decision support systems with machine learning algorithms have a potential to improve the current practices of healthcare. However, as mentioned earlier, a machine learning model cannot perfectly replicate expert's knowledge and decision making. A system without supplementary explanations/information might not be adopted in the practices. 
Our findings demonstrate that an intelligent decision support system can automatically identify salient features of decision making (e.g. rehabilitation assessment) to predict expert's decision making and generate explanations on its prediction (e.g. user-specific analysis with kinematic features). User-specific analysis from a system enables experts to gain new insights on patient's performance and reduce efforts and workload for assessment. In additions, while reviewing user-specific analysis, therapists can validate their hypothetical assessment and the correctness of a system to understand the competence of a system on a task. 
Although we demonstrate the feasibility of a decision support system for stroke rehabilitation assessment, the applied techniques of feature selection and prediction models can be utilized to other sub-domains of rehabilitation (e.g. knee rehabilitation \cite{huang2015exploring}), where an expert with limited availability needs to make a decision. 

\subsubsection{Adaptive Systems for Personalization}
During the study, we observe therapists developed different usage patterns of a decision support system and utilized different functionalities or information of a system based on a task and their own knowledge. 
Some therapists preferred to get the initial insight of assessment from feature analysis and others reviewed trajectory analysis first (Section \ref{sect:post-interviews}). Thus, making a system adaptable to each therapist's preference is recommended so that therapists can quickly collect necessary information/evidence for their decision making in practices. 


In addition, TP 6 suggested that it would be useful if a therapists can tune a system by including/excluding identified features to utilize different features based on individual therapist's experience and to correct any mismatched prediction scores. For instance, as the prediction on compensation from an interface does not predict assessment correctly time to time when a patient performs leaning compensation to the side, TP 9 commented to include a \textit{``leaning trunk to the side''} feature to predict compensation. Thus, designers should consider applying interactive techniques \cite{kulesza2015principles,lee2019intelligent} to make a system adaptive and personalized for better integration into each therapist's clinical practices.

\subsubsection{Towards Human and Machine Collaborative Systems}
In summary, instead of manually reviewing abundant features, machine intelligence can automatically identify salient features to provide useful insights on patient's performance to validate the prediction of a system and support therapist's assessment. After reviewing predicted assessment and automatically generated patient-specific analysis of a system, therapists understood the strengths and limitation of a system to consider how it can support them. Although a system sometimes disagrees with therapist's decision, therapists considered reviewing explanations is helpful to \textit{``understand patient's performance and validate my/therapist's assessment''} (TP 7). After improving understanding on patient’s performance, therapists generated more consistent evaluation (Figure \ref{fig:results-agreement}). In addition, a therapist was able to enumerate how to improve an imperfect system. 
A promising direction of future research is to explore how human and machine intelligence can complement each other to improve a complex decision making.

\subsection{Limitations}
This study aims to investigate how an intelligent decision support system with predicted assessment and user-specific analysis can support therapist's assessment on rehabilitation exercises. One of the limitations of this study is that the small sample size of participated therapists for evaluation: seven therapists from four rehabilitation centers do not represent all therapists. However, such small sample size is not unusual among similar studies \cite{hofmann2019occupational}. In addition, although therapists expressed positive opinions about an intelligent decision support system, we evaluated only one possible type of decision making of therapists, rehabilitation assessment. Other decision makings are worth exploring and require further validation. 

\section{Conclusion}
In this paper, we presented the needs of therapists during rehabilitation assessment and designed a decision support system that identifies salient features to predict the quality of motion and generate user-specific analysis as explanations on predictions. We then evaluated this system with seven therapists from four rehabilitation centers to investigate how predicted assessment and user-specific analysis of the system affect therapist's decision making on stroke rehabilitation assessment. Presenting predicted assessment and user-specific analysis increases the trust on a system and brings new insights of assessment. In addition, the proposed system enables therapists to reduce their workload (e.g. repeatedly watching videos to identify evidences) and generate more consistent assessment. Our work highlights the importance of creating user-centered and trustful machine learning-based systems to augment expert's decision making process and deploy in the practices. We believe this study can be a valuable reference to develop decision support systems for rehabilitation assessment and other critical decision supports.

\section{Appendix}
\begin{table}[h!]
\centering
\resizebox{0.5\columnwidth}{!}{%
\begin{tabular}{cccc} \toprule
\multicolumn{1}{l}{} & \multicolumn{3}{c}{\begin{tabular}[c]{@{}c@{}}Hidden Layers and Units / Learning Rate\end{tabular}} \\ \toprule
 & ROM & Smoothness & Compensation \\ \midrule
E1 & (32, 32, 32) / 0.1 & (16) / 0.0001 & (256, 256) / 0.1 \\ \midrule
E2 & (256) / 0.1 & (64, 64) / 0.001 & (128, 128) / 0.1 \\ \midrule
E3 & (256) / 0.1 & (64, 64) / 0.001 & (128, 128) / 0.1 \\ \bottomrule
\end{tabular}%
}
\caption{Parameters of Neural Networks}
\label{tab:params}
\end{table}

\bibliographystyle{ACM-Reference-Format}
\bibliography{main}

\end{document}